\documentclass[fleqn,usenatbib]{mnras}
\usepackage[utf8]{inputenc}
\usepackage{threeparttable}
\usepackage{amsmath}
\usepackage{epstopdf,color,verbatim}
\usepackage{threeparttable}
\usepackage{graphicx, color}
\usepackage{bm}
\usepackage{soul}
\usepackage{amssymb}

\usepackage{dcolumn}

\newcommand{\eqb}{\begin{eqnarray}}
\newcommand{\eqe}{\end{eqnarray}}
\newcommand{\rin}{r_{\rm i}}
\newcommand{\rout}{r_{\rm o}}
\newcommand{\bcr}{b_{\rm cr}}
\newcommand{\ah}{\alpha_{\rm h}}
\newcommand{\al}{\alpha_{\rm l}}

\graphicspath{{figures/}} 

\title[Axisymmetric plasma lenses]  
{Extreme scattering events from axisymmetric plasma lenses}

\author[]{Lingyi Dong$^1$, Maria Petropoulou$^{2,1}$, Dimitrios Giannios$^1$\\
$^{1}$ Department of Physics, Purdue University, 525 Northwestern Avenue, West Lafayette, IN, 47907, USA \\
$^{2}$ Department of Astrophysical Sciences, Princeton University, 4 Ivy Lane, Princeton, NJ, 08544, USA\\
}

\begin{document}

\date{Received.../Accepted...}

\pagerange{\pageref{firstpage}--\pageref{lastpage}} \pubyear{2018}

\maketitle

\label{firstpage}

\begin{abstract}
Frequency-dependent brightness fluctuations of radio sources, the so-called extreme scattering events (ESEs), have been observed over the last three decades. They are caused by Galactic plasma structures whose geometry and origin are still poorly understood. In this paper, we construct axisymmentric two-dimensional (2D) column density profiles for the plasma lens and explore the resulting ESEs for both point-like and extended sources. A quantity that becomes relevant is the impact parameter $b$, namely the distance of the observer's path from the lens' symmetry axis. We demonstrate its effects on the shape of ESE light curves and use it for a phenomenological classification of ESEs into  
four main types. Three of them are unique outcomes of the 2D model and do {\it not} show a characteristic U-shaped dip in the light curve, which has been traditionally used as an identification means of  ESEs.
We apply our model to five well-studied ESEs and show that elongated plasma tubes or quasi-spherical clouds are favoured over plasma sheets for four of them, while the remaining one is compatible with both lens geometries. 
\end{abstract}


\section{Introduction}\label{sec:intro}
Extreme scattering events (ESEs) are fluctuations in the brightness of radio sources  \citep{Fiedler_1987}. They typically appear as U-shaped month-long dips in the light curves of extragalactic compact radio sources and possess a strong frequency dependence \citep{Fiedler_1994}. It is generally accepted that ESEs are not related to the intrinsic source variability, but they are caused by ionized gas structures in the interstellar medium (ISM) that act as refractive lenses \citep{Fiedler_1987, Romani_1987}. 
The inferred values of the free-electron number density of the lens ($\sim 10^3-10^4$~cm$^{-3}$) suggest a pressure of at least $10^3$ times larger than the typical ISM pressure \cite[e.g.][]{Clegg_1998}. One way to overcome the over-pressure problem  is to consider a lens that is elongated along the light of sight with a length that is much larger than its other dimensions. Alternatively, the lenses could reside in regions of high pressure, such as old supernova remnants \citep{Romani_1987}. Still, the origin and  geometry of plasma lenses remain enigmatic.
 
The  free-electron column density profile of the lens is related to the lens geometry itself: a sheet of ionized plasma (planar geometry) can be described by an one-dimensional (1D) column density profile, whereas a spherical cloud or a cylindrical tube of plasma with its axis aligned to the line of sight is best described by an axisymmetric two-dimensional (2D) density profile \citep[e.g.][]{Clegg_1998, Walker_1998, Goldreich_2006, Bannister_2016, Tuntsov_2016}. 
Predictions about the properties of ESEs that are unique to each lens model may shed light into their physical origin.

In this paper, we present a detailed investigation of the 2D axisymmetric refractive lens model. We demonstrate that the shape of ESEs produced by  axisymmetric lenses depends on the so-called impact parameter, namely the perpendicular distance between the observer's path and the symmetry axis of the lens. For sufficiently large impact parameters, we discover ``atypical'' ESEs (i.e., events whose light curves do not exhibit a dip) and show that these should be more frequent at lower frequencies. We apply the axisymmetric lens model to five well sampled ESEs discovered during the monitoring program of extragalactic radio sources with the Green Bank Interferometer \citep{Fiedler_1987, Fiedler_1994}. We show that a non-zero impact parameter is crucial for describing the light curve of, at least, four events, including the high-frequency ESE of quasar 0954+658.

This paper is structured as follows. In Sect.~\ref{sec:model} we outline the 2D axisymmetric model which we then apply to five well sampled dual-frequency ESEs (Sect.~\ref{sec:apply}). We discuss the implications of our results in Sect.~\ref{sec:discuss} and conclude with a summary in Sect.~\ref{sec:summary}.
\begin{figure}
\includegraphics[width=0.49\textwidth, trim = 0 2cm 0 0]{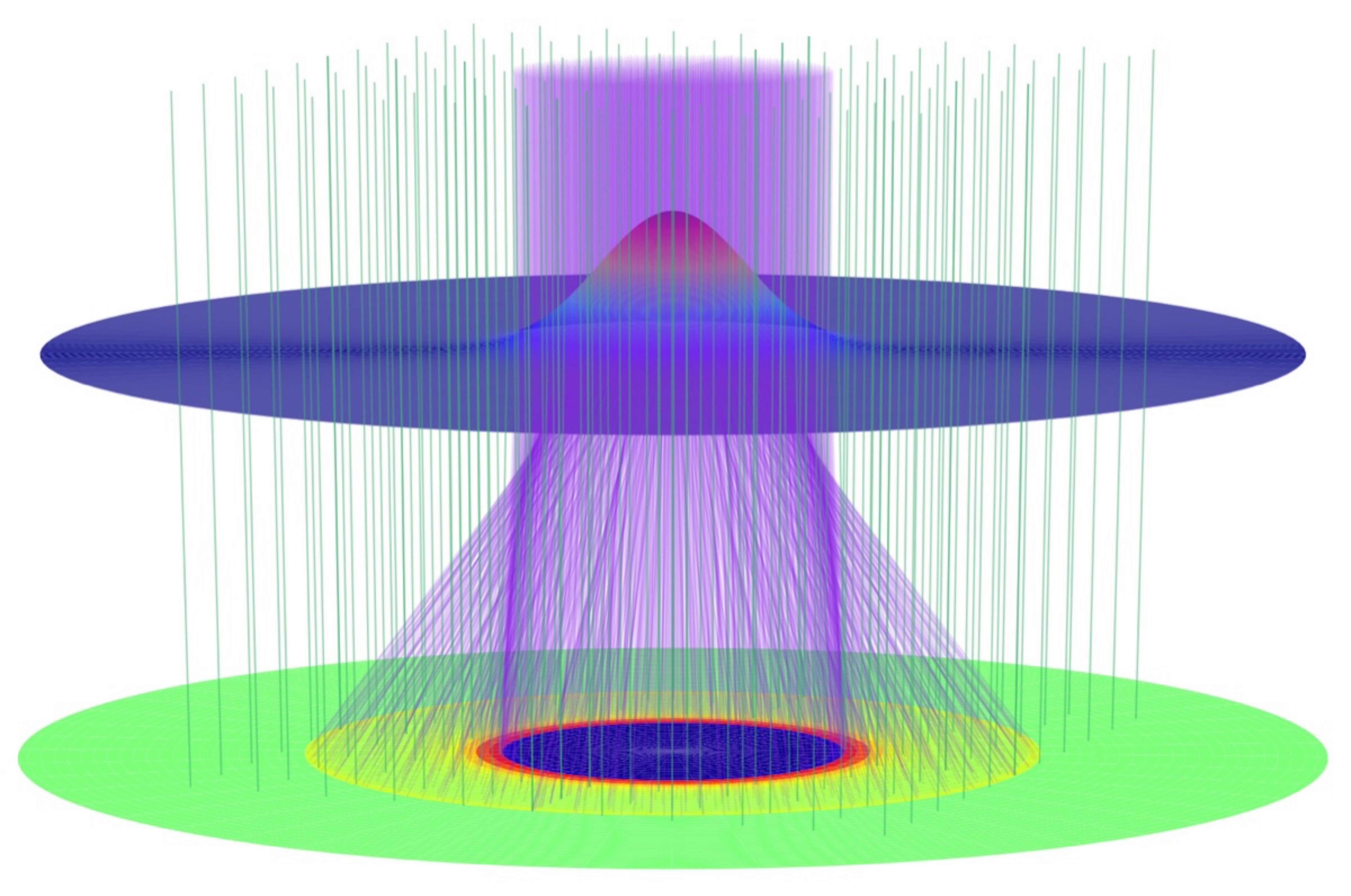}
\caption{Illustration of a refractive plasma lens with axisymmetry in the regime of geometric optics. A 3D representation of the 2D Gaussian free-electron column density is shown as a blue coloured region. Light rays from a distant source (not shown here) that are unaffected by the lens are plotted with green lines, while those that are 
refracted by more than 0.02 rad are shown in violet. A contour map of the intensity of refracted light is also shown on the observer's plane (lower plane). Two caustic rings are formed by the focusing of light rays (red coloured rings). The defocussing of light rays due to the large gradient in column density results in decreased intensity (dark blue coloured region).}  
\label{fig:lens}
\end{figure}

\section{2D axisymmetric lens model}\label{sec:model}
In the case of an axisymmetric lens, its free-electron column density $N_{\rm e}$ depends only on the distance $r^\prime$ from the symmetry axis (primed and unprimed quantities correspond to the plane of lens and the observer, respectively). We model an axisymmetric lens with a smooth 2D Gaussian profile for $N_{\rm e}$, namely:
\eqb
\label{eq:Ne}
     N_{\rm e} (r^\prime) = N_0 e^{-r^{\prime 2}/l^2}, 
\eqe
where $N_0$ is a normalization factor and $l$ is the characteristic size of the lens. The adopted density profile, which acts as a diverging lens (see Fig.~\ref{fig:lens}), may describe spherical clouds or elongated tubes of plasma overdensities \citep[see also][]{Walker_2007}.
The strength of the lens, $\alpha$, depends on frequency $\nu$ as \citep{Clegg_1998}:
\eqb 
 \alpha = \frac{q_{\rm e}^2}{\pi m_{\rm e}}\frac{N_0 D}{\nu^2 l^2} = \alpha_0 \left(\frac{\nu_0}{\nu}\right)^2, 
 \label{eq:alpha}
\eqe 
where $\nu_0=1$~GHz, unless stated otherwise,  $q_{\rm e}$ is the electron charge, $m_{\rm e}$ is the mass of the electron, $c$ is the speed of light, and $D$ is the distance from the observer to the lens\footnote{In the case of very distant sources to the lens, which are the topic of this work,  it is the distance from the observer to the lens that enters in eq.~(\ref{eq:alpha}).}.
Equation~(\ref{eq:alpha}) can also be written as $\alpha \propto \lambda \left(l_{\rm F}/l \right)^2$, where $l_{\rm F}=\sqrt{\lambda D}$ is the Fresnel scale. Diffractive effects become important, if $l < l_{\rm F}\simeq 6\times 10^{11} \text{cm} \, (\lambda/1 \, \text{m})^{1/2} (D/1 \, \text{kpc})^{1/2}$. For a lens size of $\mathcal{O}(\text{au})$ and $\nu \gtrsim 0.3$~MHz, diffraction can be safely neglected. 

The appearance of an ESE depends not only on the lens strength but also on the angular size of the lensed source, $\theta_{\rm s}$. The majority of ESEs involves background radio-loud active galactic nuclei, which appear more compact at higher radio frequencies.  We model the ratio of the angular sizes of the source and the lens as:
\eqb 
\beta_{\rm s}\equiv \frac{\theta_{\rm s}}{\theta_{\rm l}}= \beta_{\rm s0} \left(\frac{\nu_0}{\nu}\right)^{s},
\label{eq:beta}
\eqe 
where $\theta_{\rm l}\equiv l/D$. We adopt  $s=2$, unless stated otherwise \citep[see][and references therein]{Fiedler_1994, Clegg_1998}. The parameter $\beta_{\rm s}$ is crucial for the appearance of ESEs at different observing frequencies.

To compute the refracted light from a plasma lens described by eq.~(\ref{eq:Ne}) one needs to consider the radial and tangential magnification of the lensed images as well as to integrate over the azimuthal angles of incoming rays for extended sources (see  Appendix~\ref{app-a} for details). 

The lensing of a background source can result in up to three images in the observer's plane  
with an angular separation that depends on the strength of the lens and the angular extend of the source \citep[e.g.][]{Clegg_1998}.
For example, the maximum angular separation of images produced by a lens of physical size 1~AU and strength $\alpha=10$ located at 1~kpc from the observer is $\sim4.3$~mas. Such resolutions are achievable with Very-Long-Baseline Interferometry (VLBI) at $\sim$ GHz frequencies \citep[for application to ESEs, see][]{Lazio_2000, Bannister_2016}. However, not all ESEs should create multiple images resolvable with VLBI and, henceforth, we adopt the simplifying assumption that the individual images cannot be resolved.

\begin{figure}
\includegraphics[width=0.48\textwidth, trim=0 0 2cm 0]{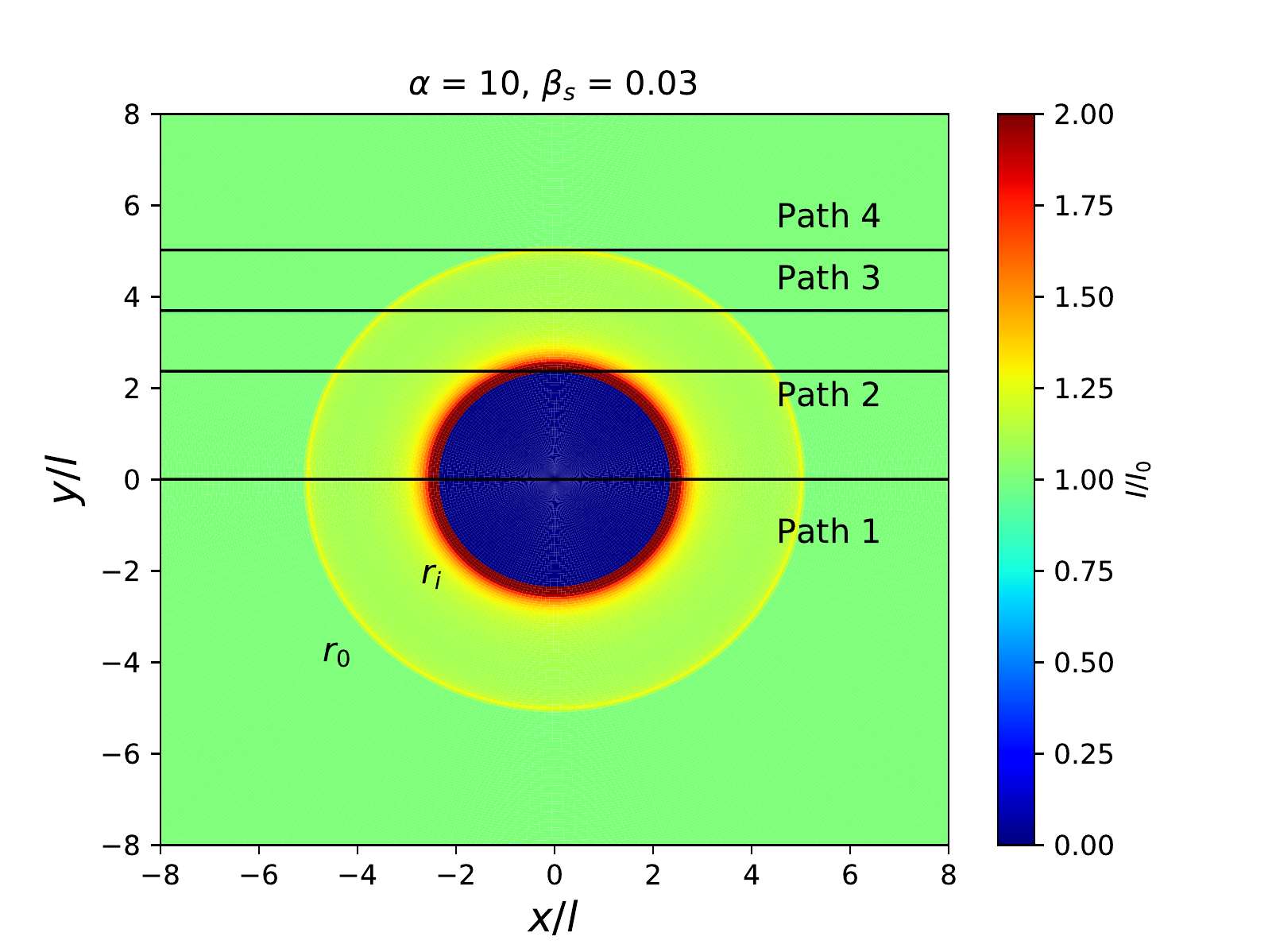} \\ 
\includegraphics[width=0.49\textwidth]{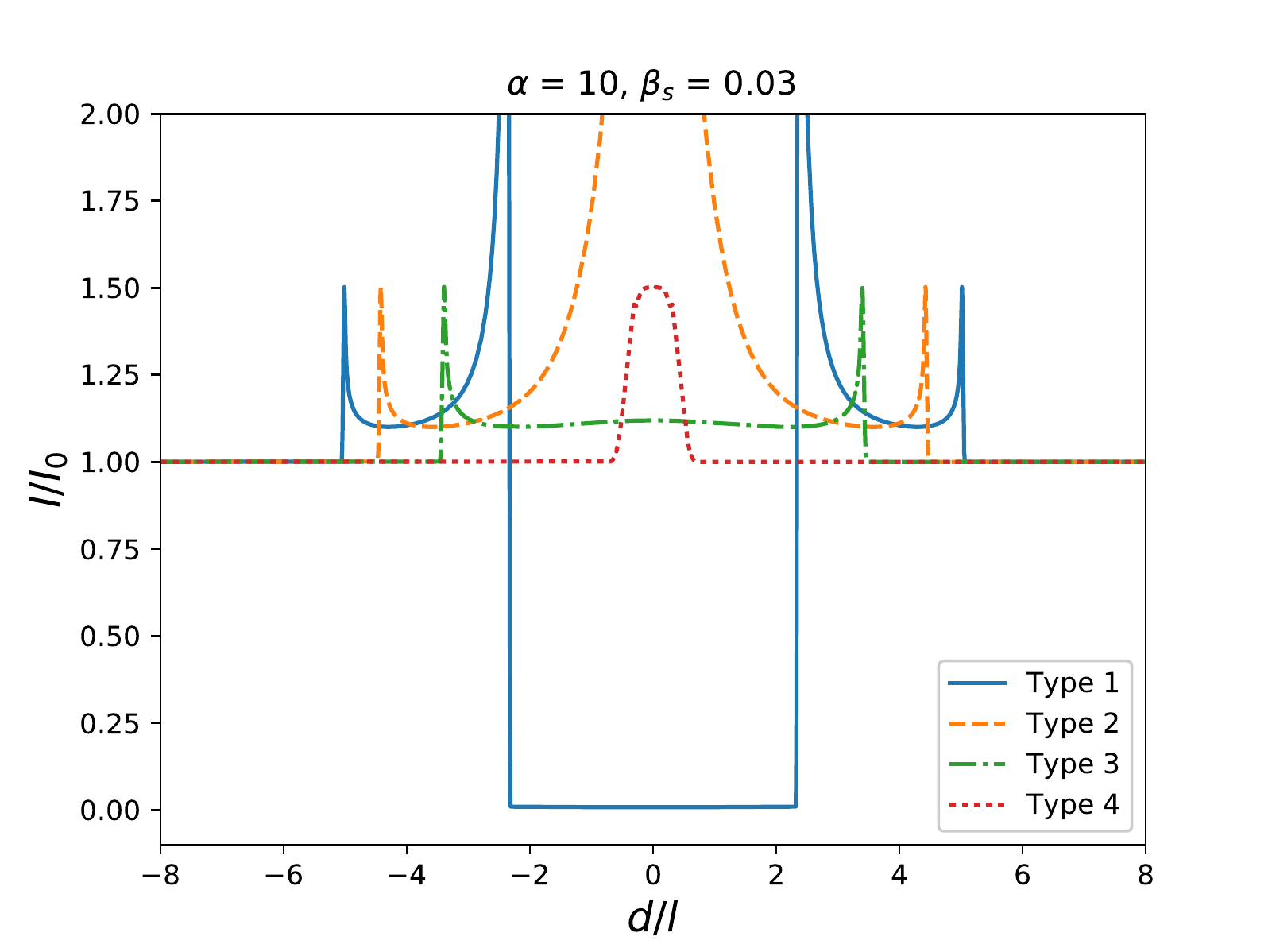}
\caption{Top panel:  Refracted intensity profile at 1 GHz of an extended radio source ($\beta_{\rm s}=0.03$) on the observer's plane caused by a 2D Gaussian lens with $\alpha=10$.  Horizontal black lines illustrate different paths of the lens with respect to a stationary observer. Each path is characterized by its ``impact parameter'' $b$, namely its perpendicular distance from the symmetry axis. Bottom panel:  Light curves of ESEs at 1 GHz obtained for the different paths shown in the top panel. Here, $d$ is the distance measured along the observer's path.}
\label{fig:2-1}
\end{figure}

The refracted intensity profile of a background source with angular extend $\beta_{\rm s}=0.03$ is shown in the top panel of Fig.~\ref{fig:2-1}. The inner circular region of minimum intensity (dip region) is caused by the refraction of light rays passing from regions of the lens with large column density gradients; the ionized plasma acts as a diverging lens. Meanwhile, the refracted intensity of the source increases at certain locations on the observer's plane (caustic rings). The number of caustic rings is a global property of the lens and depends on its strength. For example, only one caustic ring forms  for sufficiently weak lenses ($\alpha < \alpha^*=2.241$). Still, the angular extend of the background source (i.e., $\beta_{\rm s}\gtrsim 3$)  may conceal one of the caustic rings due to the convolution of the magnification factor with the source's angular profile (see Appendix \ref{app-a}).

In the case a point-like source (i.e., $\beta_s = 0$), we numerically determined
the radius of the inner ring, $\rin$,  which depends weakly on $\alpha$ and is well-described by the expression: 
\eqb 
\rin=2.02+ 0.19\ln{(\alpha - \alpha^*)}.
\label{eq:rin}
\eqe 
The radius of the outer ring can be derived analytically  \citep[see also][eq. (22)]{Clegg_1998} and is written as:
\eqb
\rout = \frac{\sqrt{2}}{2}(1+\alpha e^{-1/2})
\label{eq:rout}
\eqe 
Both radii are expressed in units of the physical lens size $l$. 

\subsection{Light curves of ESEs}\label{sec:lc}
Let us consider a lens traveling at a constant distance $D$ from the observer, which an be equivalently thought as the observer moving on a straight path with respect to the lens.
Then, the light curve of an ESE can be obtained by making a horizontal cut on the intensity profile as seen in the observer's plane. Different cuts are indicated on the plot with solid black lines (top panel in Fig.~\ref{fig:2-1}). These correspond to different paths of the observer  and $d$ denotes the distance traveled along this path  (bottom panel in Fig.~\ref{fig:2-1}).  Each path is characterized by its ``impact parameter'' $b$, namely its perpendicular distance from the symmetry axis (here, $b$ is in units of the lens size $l$). The appearance of an ESE depends strongly on $b$, as shown in the bottom panel of Fig.~\ref{fig:2-1}. Although the total flux is conserved on the image plane, it is not conserved on every individual path. Depending on the light curve shape of ESEs produced by strictly axisymmetric 2D lenses, we may classify them in the following types:
\begin{itemize}
    \item Type 1: if $b \ll \rin$, the light curve shows a dip and two spikes on each side (path 1 in Fig.~\ref{fig:2-1}). At a fixed frequency, the resulting type 1 ESEs have the same qualitative features as the ESEs produced by 1D Gaussian lenses. Any quantitative differences  (e.g., the amplitude of the spikes and dip) between the 1D lens model and the 2D lens model with $b\sim 0$ are small and arise only from the tangential magnification (see eq.~(\ref{eq:Gkt})). 
    \item Type 2:  if $b=\rin$, the light curve shows three spikes which are separated by equal time intervals  (path 2 in Fig.~\ref{fig:2-1}). This should be a rare event, as it can be realized only for a narrow range of values of the impact parameter. 
    \item Type 3: if $\rin < b <\rout$, the light curve exhibits two spikes but no dip. The lens magnifies the incident flux (path 3 in Fig.~\ref{fig:2-1}). ESEs of this type should occur more frequently than type 1 events, especially at lower frequencies, where the radius of the outer caustic ring is much larger than that of the inner caustic ring, i.e., $\rout \gg \rin$ (see also eqs.~(\ref{eq:rin}) and (\ref{eq:rout})).
    \item Type 4: if $b = \rout$, the light curve exhibits only one spike (path 4 in Fig.~\ref{fig:2-1}). Similar to type 2 events, this type of ESE should also be rare. Moreover, it should be more difficult to identify, especially if the lensed source is extended ($\beta_{\rm s} \ne 0$). In the specific example shown in Fig.~\ref{fig:2-1}, the ESE appears as a broad long-lasting low-amplitude flare.
\end{itemize} 

\subsection{Impact parameter}\label{sec:impact}
The impact parameter, introduced in the previous section, turns out to play a major role in determining the shape of observed ESEs in the 2D axisymmetric lens model. Deviations from the typical ESE (i.e., type 1 event) are expected for non-zero impact parameters.  In the following, we explore in more detail the effect of the impact parameter on ESEs.

\begin{figure}
\includegraphics[width=0.48\textwidth, trim=0 0 2cm 0]{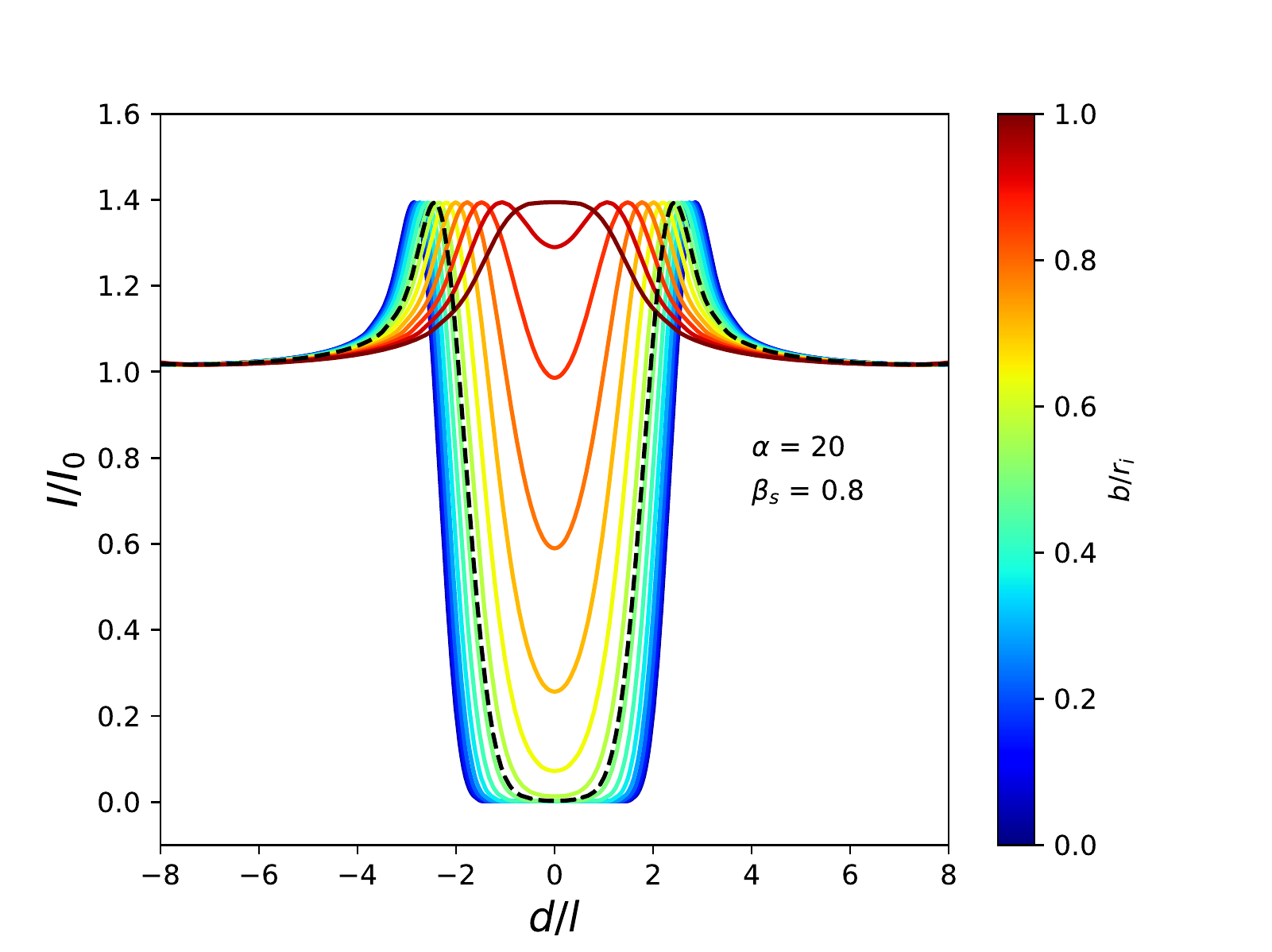} \\
\includegraphics[width=0.49\textwidth, trim=0 0 1.5cm 0]{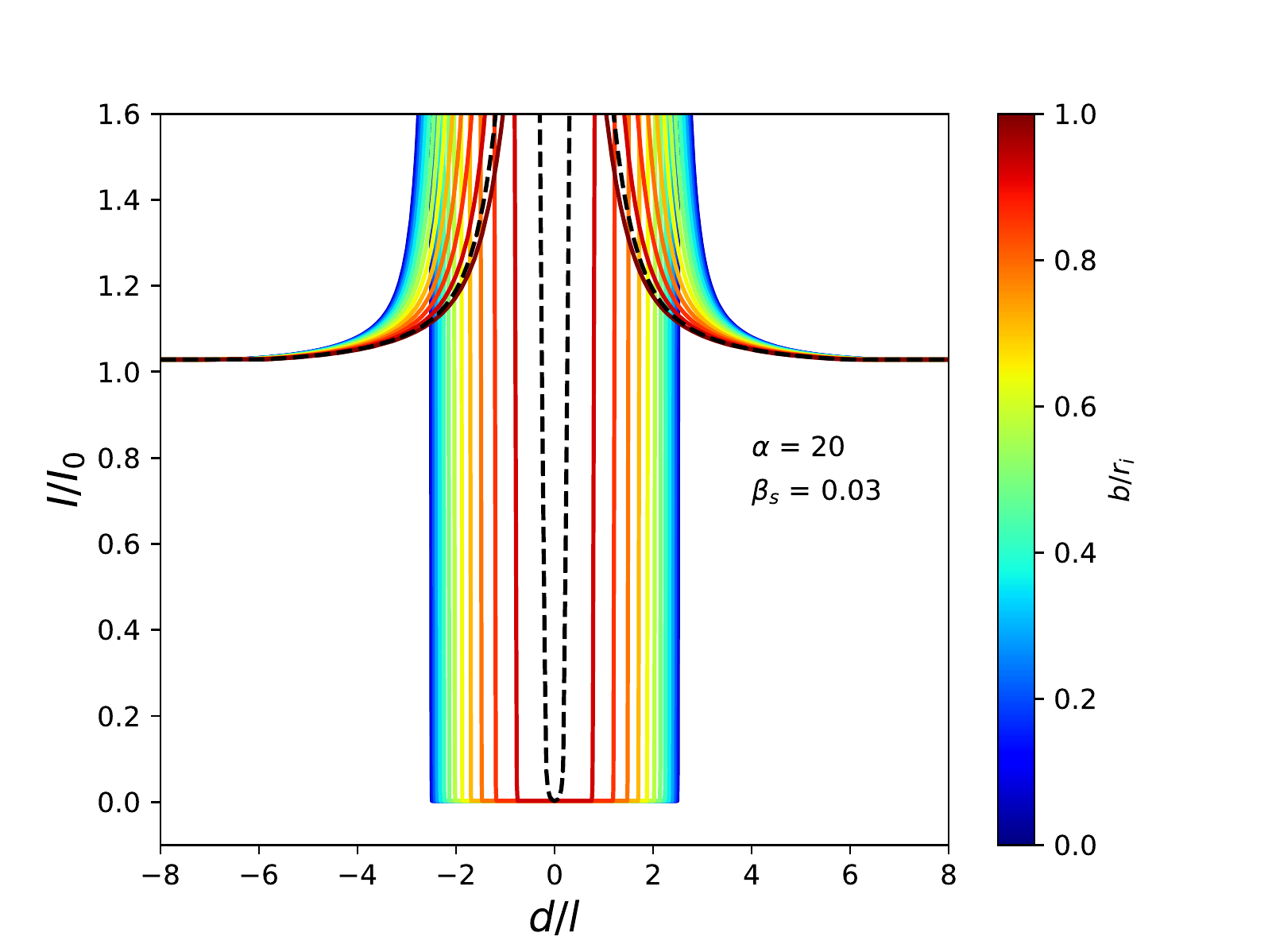}
\caption{Top Panel: Light curves of ESEs at 1 GHz caused by a 2D Gaussian lens with $\alpha= 20$, $\beta_{\rm s} = 0.8$ and different impact parameters $b$ (colour bar). The light curve obtained for $b = \bcr$ is overplotted for clarity (black dashed line). The effect of $b$ on the shape of ESEs is negligible when it is small, but it becomes important as $b\rightarrow \rin$: the dip of the light curves becomes shorter in duration, the minimum becomes rounded, and the dip-to-spike ratio becomes smaller. In the limit of $b\simeq \rin$, the two spikes merge into one and the dip disappears.  Bottom panel: Same as in the top panel, but for a point-like source with $\beta_{\rm s} = 0.03$.  }
\label{fig:lc_b}
\end{figure}

Figure~\ref{fig:lc_b} shows the dependence of the ESE light curves (at $\nu=1$~GHz) on the ratio $b/\rin$ (colour bar), for an extended source (top panel) and a point-like source (bottom panel). In both cases, we find that the ESE light curves produced by a 2D Gaussian lens with $b=0$ are very similar to those produced by an 1D lens. However, for $b > 0$ the shapes of ESEs from a 2D lens begin to deviate significantly from those produced by an 1D lens.  The flattened shape of the dip is preserved as long as $b$ is smaller than a critical value, $\bcr$. The latter is larger for point-like sources; we find $\bcr \simeq 0.98\, \rin$ for $\beta_{\rm s0}=0.03$ (bottom panel) and $\bcr\simeq 0.5 \, \rin$ for $\beta_{\rm s0}=0.8$ (top panel).  For $b > \bcr$, the light curve obtains a rounded shallow minimum and the two spikes get closer while retaining their amplitudes. A 2D axisymmetric lens with $b> \bcr$ can, therefore, explain ESEs with a shallow dip and prominent spikes, even for non point-like sources.  
On the contrary, a shallow dip in the context of the 1D Gaussian model requires either a weak lens or an extended source \citep[e.g.][]{Clegg_1998}, both of which result in spikes with small amplitude (see next section).  

\begin{figure}
 \centering 
 \includegraphics[width=0.48\textwidth, trim=0 0cm 2cm 0]{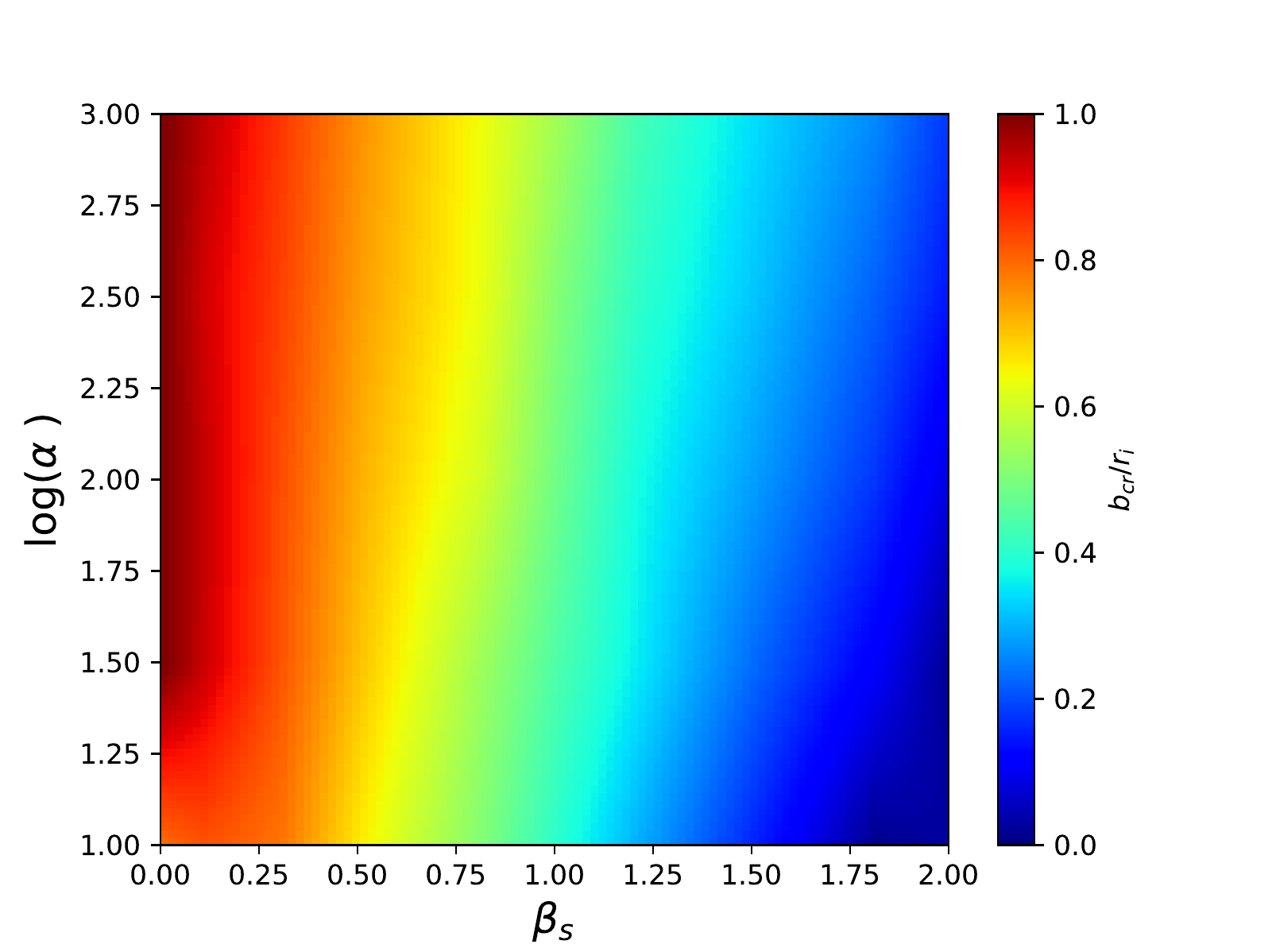}
 \caption{Colour map of the ratio $\bcr/\rin$ as a function of $\alpha$ (in logarithmic scale) and $\beta_{\rm s}$ (in linear scale). The ratio is insensitive on the former, but it is almost inverse proportional to the latter.} 
 \label{fig:bcr}
\end{figure} 
We numerically determined the critical impact parameter for different values of $\alpha$ and $\beta_{\rm s}$ in the ranges $[10, 1000]$ and $[0,2]$, respectively.  We find that $\bcr$ is only weakly dependent on the strength of the lens, whereas it scales as $\bcr \propto 1/\beta_{\rm s}$ -- see Fig.~\ref{fig:bcr}.
Fig.~\ref{fig:bcr} shows that for very extended sources (i.e., $\beta_{\rm s} \gtrsim 2$) the critical impact parameter approaches zero. In this regime and for $b\le \rin$, ESEs produced by both 1D and 2D Gaussian lenses will have light curves with rounded dips. Nevertheless, atypical ESE light curves (in the context of 1D Gaussian lenses) can still be obtained for $b\ge \rin$. 
\begin{figure}
 \centering 
 \includegraphics[width=0.48\textwidth, trim=0 0cm 2cm 0]{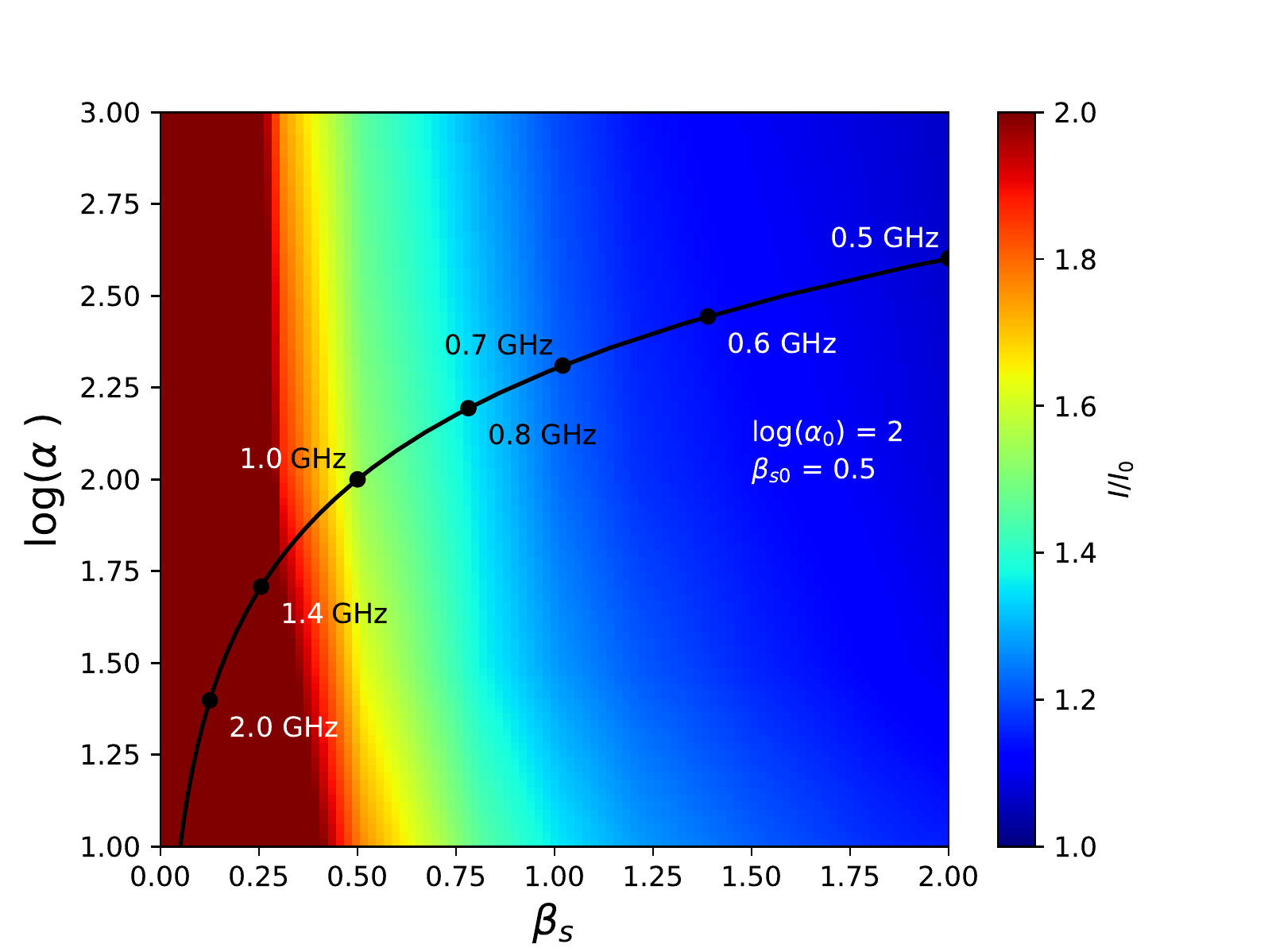}
 \caption{Colour map of the ratio of lensed-to-unlensed brightness $I/I_0$ at the inner caustic as a function of lens strength log$(\alpha)$ and source size $\beta_{\rm s}$ for $b = 0$. Overplotted is the dependence of log$(\alpha)$ and $\beta_{\rm s}$ with frequency (black line). For clarity, seven points with the frequency information are marked along the curve. At low frequencies ( $< 1$ GHz) the ratio decreases, although the lens becomes stronger ($\alpha \propto \nu^{-2}$). This is due to the larger extent of the background source at low frequencies, $\beta_{\rm s} \propto \nu^{-2}$, which smooths the light curve.}
 \label{fig:inten_cm}
\end{figure}

The detectability of ESEs also depends on the ratio of the lensed-to-unlensed radio brightness (i.e., $I/I_0$). To demonstrate its dependence on frequency, we computed  $I/I_0$ at the inner caustic of an ESE produced by a 2D Gaussian lens for different values of $\alpha$ and $\beta_{\rm s}$, and for $b = 0$ -- see Fig.~\ref{fig:inten_cm}. As the ratio of intensities at the inner caustic is not affected by the impact parameter, the results displayed in Fig.~\ref{fig:inten_cm} are applicable to $b > 0$, too. Although the lens becomes stronger at lower frequencies (see eq.~(\ref{eq:alpha})), the relative intensity is typically lower than that at higher frequencies. The main reason for this is that the extension of the background source becomes larger at lower frequencies (see eq.~(\ref{eq:beta})). In general, large values of $\beta_{\rm s}$ tend to smoothen  the light curves.  

Our analysis suggests that a large sample of ESEs observed at multiple frequencies would be ideal for identifying cases explained only by axisymmetric plasma lenses. 
 
\section{Application to observations}\label{sec:apply}
We apply the 2D Gaussian lens model to five ESEs discovered during the monitoring program of extragalactic radio sources with the Green Bank Interferometer \citep{Fiedler_1987, Fiedler_1994}. The selected ESEs have good temporal coverage (i.e., no gaps) at both observing frequencies (2.25 GHz and 8.3~GHz). 

Theoretical light curves of ESEs were computed in terms of the dimensionless ratio $d / l$ (e.g., Fig.~\ref{fig:lc_b}), where $d$ is the distance measured along the observer's path and $l$ is the physical size of the lens (i.e., its radius for axisymmetry). Assuming that the relative transverse velocity $v$ is constant, the ratio $d /l $ can be associated with the duration of an ESE, $\Delta t$, as follows:
\begin{equation}
   v= \frac{d}{\Delta t} = \frac{l}{\tau}
\end{equation}
where $\tau$ is some scaling factor (in units of time), determined by matching the observed to the theoretical ESE light curve (see Table~\ref{tab:tab2}).   

While fitting the observed ESEs (Sect.~\ref{sec:apply}), we introduced another free parameter to the model, i.e., the fraction of the source intensity that is being lensed. In cases where the brightness profile of the background source remained constant in time, the  flux of the lensed (or unlensed) component corresponds to a constant  value. It can be also expressed as a constant fraction of the source intensity, if the latter is changing with time (see Table~\ref{tab:tab2}).  

In all cases, we assumed a lens located at $D=1$~kpc with a size $l=\left(\theta_{\rm s0}/\beta_{\rm s0}\right)D$, where $\theta_{\rm s0}$ is the angular source size at 1 GHz. The latter was estimated using published results of the angular size at other frequencies (typically, at 5 GHz) and assuming a scaling law $\theta_{\rm s}\propto \nu^{-2}$, as in eq.~ (\ref{eq:beta}). We discuss each case separately in the following paragraphs. Our fitting results are summarized in Table~\ref{tab:tab2} at the end of this section. 

\subsection{Interesting cases}
\subsubsection{Q0954+658}
\begin{figure}
 \centering 
 \includegraphics[width=0.49\textwidth]{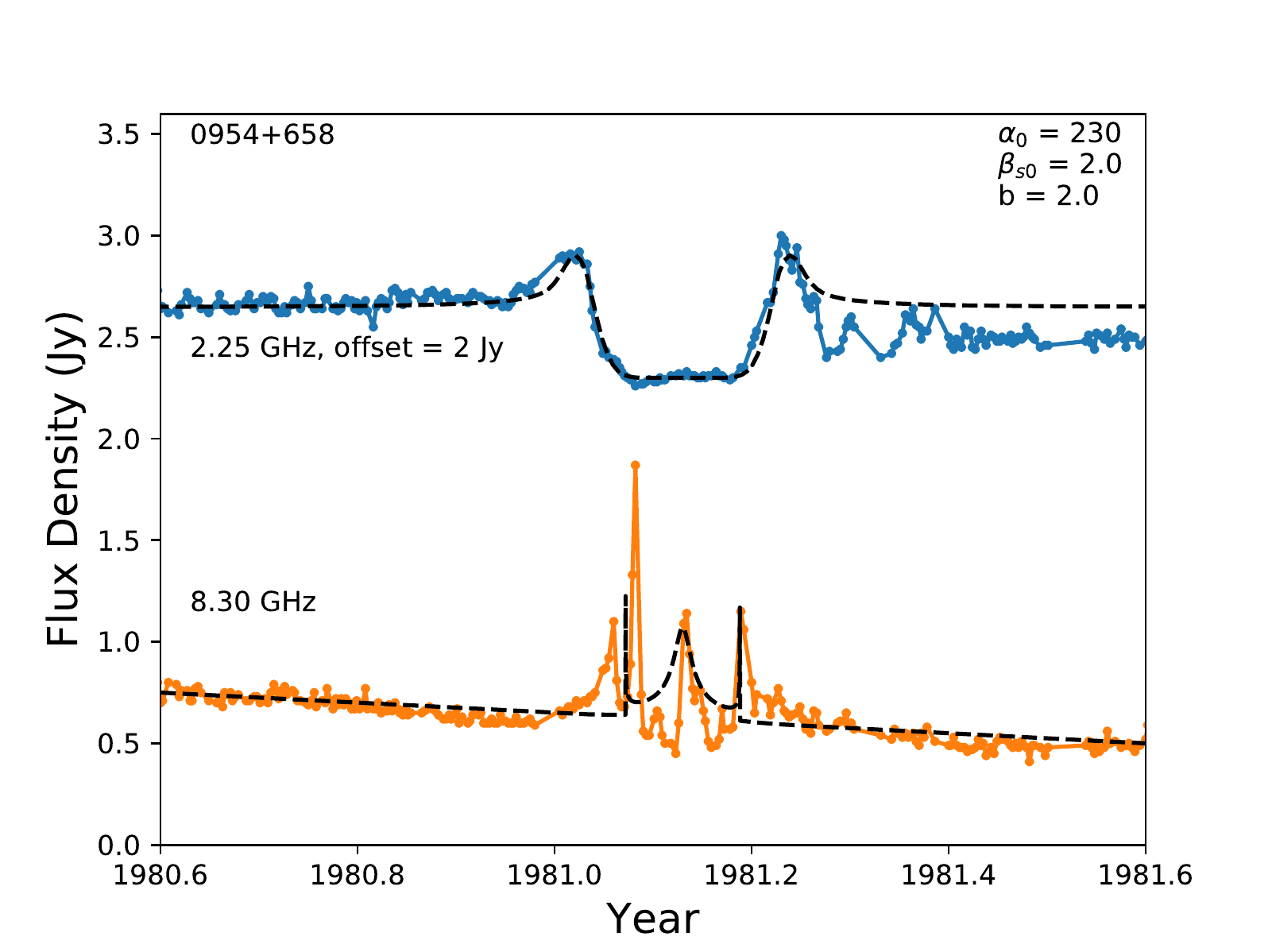} 
 \caption{Light curves of quasar 0954+658 at 8.3 GHz (orange coloured symbols) and 2.25 GHz (blue coloured symbols) focused around the period of the ESE. Best-fit light curves obtained for an axisymmetric lens with $\alpha_0=230$, $\beta_{\rm s0}=2.0$, and $b=2.0$ are overplotted (black dashed lines). The 2.25 GHz light curve is plotted with an offset of 2 Jy. }
 \label{fig:0954+658} 
\end{figure}

The ESE with the best coverage, so far, is the one detected in the light curve of quasar 0954+658 at the beginning of 1981. It appears as a typical ESE at low frequencies (2.25 GHz), but it has an irregular shape with several spikes at 8.30 GHz (see Fig.~\ref{fig:0954+658}). The source flux at that frequency was decreasing at a rate 0.25 Jy/yr during 1981. The dual-frequency light curve of the ESE poses a challenge to the 1D Gaussian lens model, as this fails to reproduce the high-frequency ESE \citep[see also][]{Walker_1998}. 

We argue that the ESE of 0954+658 at 8.3 GHz can be interpreted as a type 2 event caused by an axisymmetric lens, for the following reasons:
\begin{itemize}
 \item it is a rare event; to the best of our knowledge, no other similar ESE has been detected so far. 
 \item it resembles the 3-spike structure of a type 2 event caused by a smooth 2D Gaussian lens (see bottom panel in Fig.~\ref{fig:2-1}); any inhomogeneities on top of the smooth column density profile could explain the two spikes with very short temporal separation and different peak fluxes that were observed at the beginning of the ESE.
\end{itemize}
By interpreting the high-frequency ESE of  0954+658 as a type 2 event we can constrain the strength of the axisymmetric lens as follows. Let $\ah$ and $\al$ be the strength of the lens at $\nu_{\rm h}=8.3$~GHz and $\nu_{\rm l}=2.25$~GHz, respectively.  The appearance of a type 2 event at $\nu_{\rm h}$ requires that $b=\rin(\ah)$. The ratio of durations of the ESE at the two frequencies can be written as:
\eqb 
\frac{\Delta t_{\rm h}}{\Delta t_{\rm l}} = \sqrt{\frac{\rout^2(\ah)-\rin^2(\ah)}{\rin^2(\al)-\rin^2(\ah)}},
\label{eq:ratio}
\eqe 
where we also used the fact that the ESE appears as a typical type 1 event at $\nu_{\rm l}$. Since $\Delta t_{\rm h}\simeq \Delta t_{\rm l}$, we obtain the following constraint, $\rout(\ah) \simeq \rin(\al)$, which results in  $\alpha_0\simeq 230$ when combined with eqs.~(\ref{eq:alpha}), (\ref{eq:rin}), and (\ref{eq:rout}). We can then infer the impact parameter as $b=\rin(\ah)\simeq 2$. The expressions used to infer the lens properties are exact only for point sources. Still, for sufficiently compact sources ($\beta_{\rm s0}<1$), the analytical estimates are close to those inferred from the actual fit to the ESEs.

Our best-fit results are presented in Fig.~\ref{fig:0954+658}. A strictly axisymmetric lens model can capture the basic features of the ESE at both frequencies. 
The mismatch between the model and the data at 8.3~GHz can be attributed to inhomogeneities of the free-electron column density, which in our model was described by a smooth and well-behaved function. Alternatively, a small amount of shear may produce its own double spiked magnification patterns, which cannot be accounted for in the idealized case of a pure axisymmetric lens. 
Although the requirement for axisymmetry in the case of 0954+658 has been already pointed out in earlier studies \citep{Walker_1998, Walker_2001, Walker_2007}, our interpretation of the spikes at the 8.3 GHz light curve is different. 

\begin{figure}
 \centering 
 \includegraphics[width=0.49\textwidth]{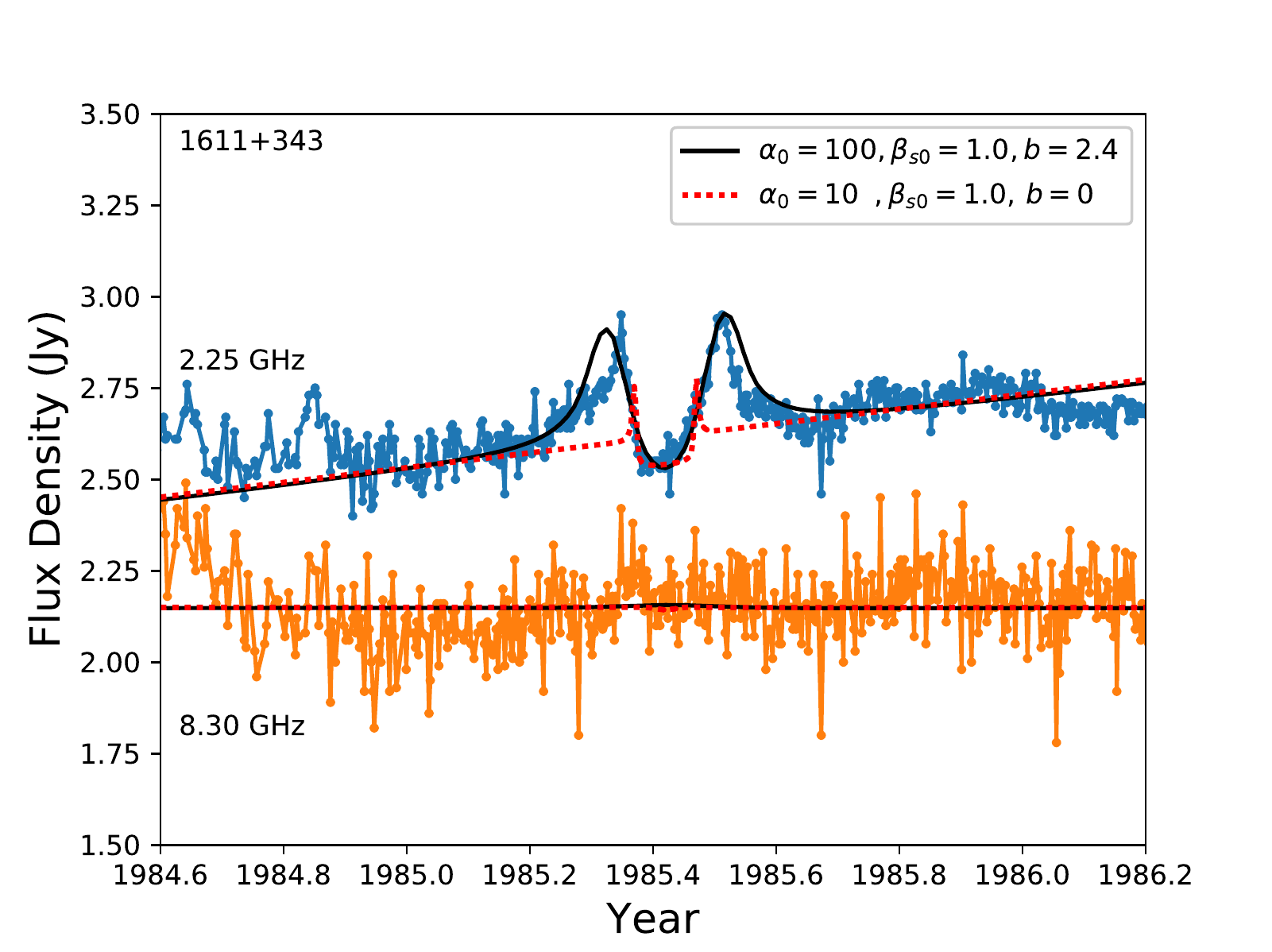} 
 \caption{Light curves of quasar 1611+343 at 8.3 GHz (orange coloured symbols) and 2.25 GHz (blue coloured symbols) focused around the period of the ESE. Red dotted lines: The best-fit light curves obtained from a 2D Gaussian lens, assuming $b = 0$. The dip of the ESE is fitted well, but the model fails to explain the spikes before and after the dip. Black solid lines: The best-fit light curves obtained assuming an axisymmetric 2D lens with $b \ne 0$. The parameters inferred from the fit are: $\alpha_0=100$, $\beta_{\rm s0}=1.0$, and $b=2.4$. Here, we assumed that $9\%$ of the total source flux is lensed.   } 
 \label{fig:1611+343}
\end{figure} 

\subsubsection{Q1611+343}
The ESE of quasar 1611+343 occurred in 1985 and is another example that supports the axisymmetric lens model. The low- and high-frequency light curves  of the source are presented in Fig. \ref{fig:1611+343}. At 2.25 GHz the source flux was increasing at a rate 0.2 Jy/yr during 1985.  

What makes this event particular is the appearance of the ESE at 2.25 GHz: a shallow dip with two prominent spikes. In other words, the light rays that are being refracted out to produce the dip are fewer than those accumulated at the caustics to produce the spikes in the light curve. In the framework of the 1D lens model or the 2D lens model with $b\rightarrow 0$, a shallow dip in the ESE light curve suggests a weak lens or an extended background source. This is illustrated in Fig.~\ref{fig:1611+343}, where the ESE produced by a 2D lens with $b=0$, $\alpha_0=10$, and $\beta_{\rm s0}=1$ is plotted with red dashed line. No other combination of parameters can reproduce the low-frequency light curve, unless $b>0$.  The data suggest a large impact parameter, i.e. $b > \bcr$, as discussed in Sect.~\ref{sec:impact}. The resulting light curve is shown in the bottom panel of \ref{fig:1611+343}. 

\begin{figure}
 \centering 
 \includegraphics[width=0.49\textwidth]{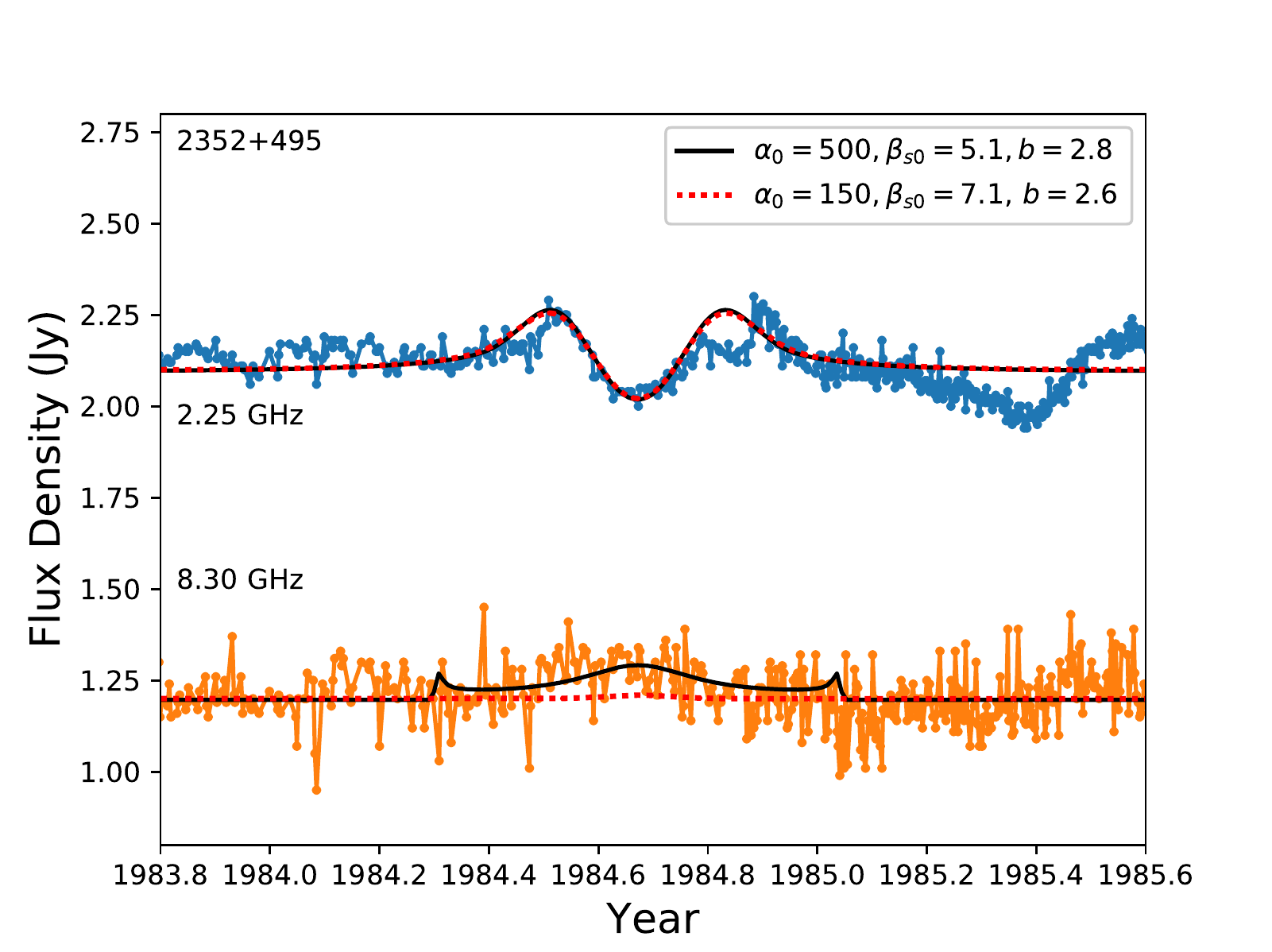} 
 \caption{Light curves of quasar 2352+495 at 8.3 GHz (orange coloured symbols) and 2.25 GHz (blue coloured symbols) focused around the period of the ESE. 
Red dotted lines: The best-fit light curves obtained without trying to model the small flux increase observed at 8.3 GHz. Black solid lines: The best-fit light curves obtained when the small flux increase at 8.3 GHz is taken into account. Although both parameter sets (see inset legend) have similar non-zero impact parameters and can describe the event at 2.25 GHz equally well, the lens strengths and angular source sizes are very different. 
}
 \label{fig:2352+495}
\end{figure}
\subsubsection{Q2352+495}
The 2.25 GHz light curve of quasar 2352+495 during 1984-1986 shows two ESEs \citep[Fig.~1 in][]{Fiedler_1994}. Here, we focus on the one that occurred during 1984, as this was accompanied by a period of increased flux at 8.3 GHz. We find that the ESE at 2.25 GHz can be equally well described by two parameter sets with similar non-zero impact parameter but very different lens strengths and angular source sizes (see inset legend in Fig.~\ref{fig:2352+495}). Interestingly, for $\alpha_0=500$, $\beta_{s0}=5.1$, and $b=2.8$ the ESE at 8.3~GHz resembles that of a type 2 event (see Fig.~\ref{fig:2-1}), where the central spike appears very broad and smooth because of the large angular source size. We refer to this set of parameters as 2352+495(a), and refer to the other set as 2352+495(b).  As it is not possible to tell if the increased flux at 8.3~GHz is intrinsic to the source or a result of an ESE, we cannot lift the degeneracy between the two models considered here. Future detections of ESEs at more than two frequencies are crucial for constraining the properties of plasma lenses \citep[see also][]{Bannister_2016}.
 
\subsection{Other cases}
\begin{figure}
\centering
\includegraphics[width=0.49\textwidth]{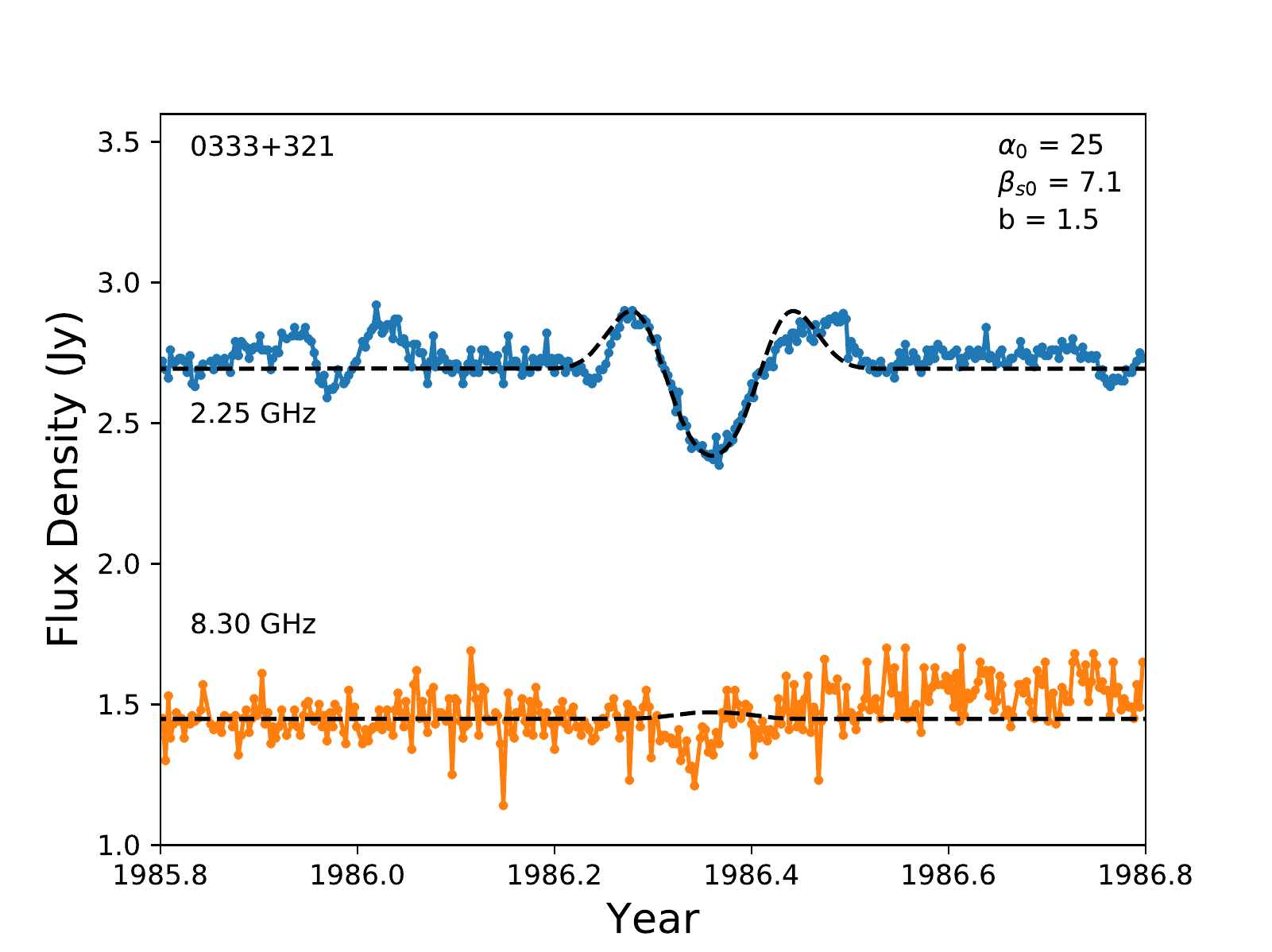} 
\caption{Same as in Fig.~\ref{fig:2352+495} for $\alpha_0=25$, $\beta_{\rm s0}=7.1$, and $b=1.5$.
}
\label{fig:other1}
\end{figure}

\begin{figure}
\centering
\includegraphics[width=0.49\textwidth]{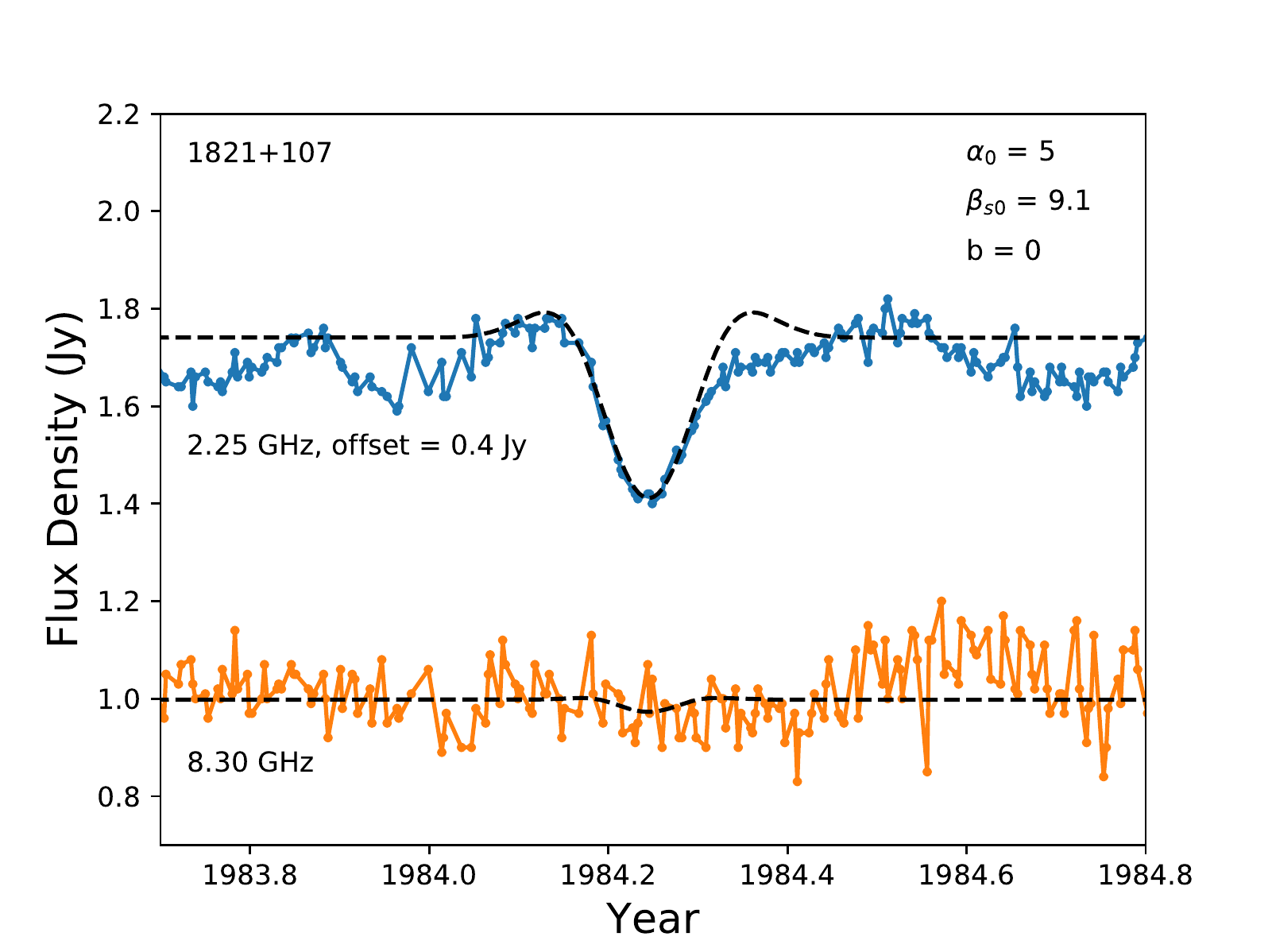} 
\caption{Same as in Fig.~\ref{fig:2352+495} for $\alpha_0=5$, $\beta_{\rm s0}=9.1$, and $b=0$.}
\label{fig:other2}
\end{figure}

The ESEs detected in the light curves of quasars 0333+321 and 1821+107 are type 1 events according to our classification scheme (see Sect.~\ref{sec:model}). They can be fitted by an axisymmetric lens model with $b < \rin$ (see Figs.~\ref{fig:other1}-\ref{fig:other2}) or equivalently by an 1D lens model \citep[see also][for PKS 1939--315]{Bannister_2016}.

\begin{table*}
\centering 
\caption{Parameter values of the 2D Gaussian lens model for five ESEs. Below $\tau$ represents the scaling factor obtained by comparing theoretical and observed light curves and is inversely proportional to the speed of the lens. ``Lensed'' and ``Unlensed'' denote, respectively, the component of the source that has been lensed by the plasma lens or not (for more details, see Sect.~\ref{sec:apply}). ``L'' and ``H'' stand for  low (2.25 GHz) and high (8.3 GHz) frequencies, respectively.}
\label{tab:tab2}
\begin{tabular}{ccccccccccc}
 \hline 
 Source & $\alpha_0$ & $\beta_{\rm s0}$ & $b$ & $s$ & $\tau$(yr) & Date & Lensed (L) & Unlensed (L) & Lensed (H) & Unlensed (H)\\
 \hline 
 0954+658 &  230 & 2.0 & 2.0 & 2 &  0.05 & 1981.1 & 0.35 Jy& 0.3 Jy & 7\%& 93\% \\
 1611+343 &  100 & 1.0 & 2.4 & 2 & 0.11 & 1985.4 & 9\% & 91\% & 0.2 Jy  & 1.95 Jy  \\
 2352+495 (a) & 500 & 5.1 & 2.8 &  2 & 0.09 & 1984.7 & 0.7 Jy & 1.4 Jy & 0.3 Jy & 0.9 Jy  \\
 2352+495 (b) & 150 & 7.1 & 2.6 &  2 & 0.08 & 1984.7 & 0.8 Jy & 1.3 Jy & 0.3 Jy & 0.9 Jy  \\
 0333+321 & 25 & 7.1 & 1.5 & 2 & 0.04 & 1986.4 & 0.6 Jy & 2.1 Jy & 0.4 Jy &  1.05 Jy \\
 1821+107 & 5 & 9.1 & 0 & 2 & 0.05 & 1984.2 & 0.85 Jy & 0.5 Jy & 0.4 Jy & 0.6 Jy \\
 \hline
\end{tabular}
\end{table*}

\begin{table*}
\centering 
\caption{Physical properties of the lenses as obtained from the parameter values shown in Table~\ref{tab:tab2}.  In all cases, we assumed a spherical lens geometry, a plasma temperature $T=10^4$~K, and a distance $D=1$~kpc. The angular sizes of sources are adopted by \citet{Gabuzda_1996} and \citet{Fey_1996}; these are extrapolated to 1 GHz using the scaling $\theta_{\rm s}\propto \nu^{-2}$. }
\label{tab:tab3}
\begin{tabular}{cccccc}
 \hline 
 Source & $\theta_{\rm s0}$ [mas] & $l$ [cm] & $v$ [km s$^{-1}$] & $N_0$ [cm$^{-2}$] & $p_{\rm e}$ [K cm$^{-3}$] \\
 \hline 
 0954+658 & $\sim 7$  & $5.2\times 10^{13}$ & $3.3\times 10^2$ & $2.7\times 10^{18}$ &  $5.1\times 10^8$\\
 1611+343 & $\sim 11$  & $1.6\times 10^{14}$ & $4.7\times 10^2$ & $1.2\times 10^{19}$ & $7.0\times 10^8$   \\
 2352+495 (a)  & $\sim 6$ &  $1.8\times 10^{13}$ & $6.2\times 10$ & $6.6\times 10^{17}$ &  $3.7\times 10^8$ \\
 2352+495 (b) & $\sim 6$ &  $1.3\times 10^{13}$ & $4.5\times 10$ & $1.0\times 10^{17}$ &  $8.1\times 10^7$ \\
 0333+321 & $\sim 6$ & $1.3\times 10^{13}$ & $1.0\times 10^2$ & $1.7\times10^{16}$ & $1.3\times 10^7$  \\
 1821+107 & $\sim 3$  & $4.9\times 10^{12}$ & $3.1\times 10$ & $5.2\times10^{14}$ & $1.0\times 10^6$ \\
 \hline
\end{tabular}
\end{table*}

\section{Summary and Discussion}\label{sec:discuss}
In this paper, we investigated the observational properties of ESEs caused by lenses with 2D pure axisymmetric column density profiles in the limit of geometric optics. In addition to the refractive strength of the lens ($\alpha$) and the relative angular size of the source to the lens ($\beta_{\rm s}$), we introduced a new parameter, the so-called impact parameter $b$, which denotes the perpendicular distance between the path of the observer and the symmetry axis of the lens.

Although the number of caustic rings depends solely on the global properties of the lens, we demonstrated that the appearance of ESE light curves produced by a 2D Gaussian lens depends sensitively on the impact parameter. We classified the resulting ESEs into four types, depending on the relation of $b$ with the radii of the inner and outer rings. All types of events, but one (type 1), are unique products of the 2D axisymmetric lens model. In particular, ESEs with three spikes (type 2) are expected when $b=\rin$. The symmetric pair of spikes is more prominent for point-like sources and, thereby, the appearance of type 2 events strongly depends on observing frequency. Regardless, these should be very rare events. For $b>\rin$, any magnification observed in the light curves (type 3 and 4 events) will be a result of the outer caustic ring crossing. The spikes in the type 3 and 4 ESE light curves are more pronounced  for stronger lenses and/or background sources with smaller angular extent. In addition, the duration of the spikes is typically short. All the above  suggests that type 3 and 4 ESEs might be harder to detect. Nevertheless, they should be more frequent at lower observing frequencies (see text in Sect.~\ref{sec:lc}). Thus, low-frequency surveys, such as the  Murchison Widefield Array \citep{tingay_13}  and the  Australian Square Kilometre Array Pathfinder \citep[ASKAP,][]{johnston_08} with good temporal resolution may reveal more of the atypical ESEs. 
The Canadian Hydrogen Intensity Mapping Experiment \citep[CHIME,][]{CHIME_2018} operating at 400-800 MHz may also discover ESEs while surveying for pulsars and fast radio bursts. However, being a non-imaging instrument, it may not be very efficient at monitoring slowly evolving sources, like radio-loud active galaxies. At GHz frequencies, MeerKAT \citep{meerkat1}, with smaller field-of-view but higher sensitivity than ASKAP, will also be efficient at detecting ESEs.

The details of the U-shaped dip of an ESE can also be used to probe the lens geometry. We showed that in the 2D Gaussian lens model there is a critical value of the impact parameter, $\bcr$, beyond which a dip with rounded bottom and a smaller dip-to-spike ratio can be produced. We numerically determined the ratio $\bcr/\rin$ for a wide range of lens' strengths $\alpha$ and source sizes $\beta_{\rm s}$ (Fig.~\ref{fig:bcr}). The detection of an ESE of a point-like source (i.e., small $\beta_{\rm s}$) with rounded bottom cannot be explained by the 1D model, and is indicative of a non-zero impact factor (see Fig.~\ref{fig:1611+343}). However, for extended sources, ESEs with rounded dips can be explained, in general, by both 2D and 1D lenses. So, we expect more atypical ESEs (e.g., type 2 and those with non-zero impact factor) when the sources are more compact and the lenses are stronger.

We applied the 2D lens model to five well sampled ESEs and showed that four  of them support the scenario of axisymmetric lenses (cylindrical tubes or spheres). The remaining one can be explained by either axisymmetric or planar geometries.
The size of lens can also be determined as $l = D (\theta_{\rm s0}/\beta_{\rm s0})$, where  we assumed a distance of $D=1$~kpc and extrapolated published values of  $\theta_{\rm s}$ to 1~GHz assuming a $\nu^{-2}$ scaling. The free-electron column density $N_0$ can be then derived using eq.~(\ref{eq:alpha}) and the values of $\alpha_0$ and $l$  (see Table~\ref{tab:tab2}).  If the axisymmetric model describes spherical lenses, we may estimate the free-electron pressure as $p_{\rm e}/k=N_0 T/l$ where $T$ is the electron temperature. Our results are summarized in Table~\ref{tab:tab3}. In all cases, the inferred pressure exceeds that of the ISM by many orders of magnitude, in agreement with past studies (e.g., \cite{Bannister_2016, Clegg_1998}). The constraints on the free-electron density can be relaxed, if the lenses are cylindrical tubes with one dimension being much larger than $l$. The geometry of the lens may also be related to the formation mechanism and, in turn, the formation site. \cite{Fiedler_1994} demonstrated that the sources with ESEs have a small angular separation from the Galactic radio continuum loops (with $\sim 1\%$ chance coincidence probability). They concluded that the Galactic loops may provide the necessary conditions for the formation of plasma lenses. 
Other authors also considered how does the elongated plasma structure along the line of sight solve the overpressure problem we encountered above. For example, \cite{Pen_2014} and \cite{Simard_2018} proposed elongated and corrugated plasma sheets along the line of sight as means of explaining pulsar scintillation and ESEs.

Recently, \cite{Tuntsov_2016}  presented a technique for fitting ESEs that does not rely on a presumed smooth column density profile. The column density can be, instead, reconstructed from the dynamic spectrum interpolated from observations at multiple frequencies \citep{Bannister_2016}. 
However, their method can only be applied to point sources, thus limiting its applicability. On the contrary, our method, being an  extension of the methodology presented by \cite{Clegg_1998}, can be applied to extended sources and at all frequencies (within the limit of geometric optics. 
Our method is also computationally fast, because the refracted intensity profile on the observer's plane is also axisymmetric. For point sources, it can be as fast as the calculation for 1D lens; in the extended source case, it requires the integration of the azimuthal component of the source (see eq.~(\ref{eq:intensity})), which only increases the computational time by a factor of ten.   This makes it ideal for the analysis of large data sets and at frequencies below 1 GHz. Yet, our method is limited to the idealized case of an axisymmetric lens with a specific and well behaved column density profile, and thus cannot capture any finer structures present in ESEs (see e.g. Fig.~\ref{fig:0954+658}).

\section{Conclusion}\label{sec:summary}
The unusual U-shaped dips in the luminosity of radio sources have been traditionally used as an identification means of  ESEs. However, if the lenses are predominantly plasma structures with  axisymmetric column density profiles, then 
more ESEs with shapes different than those produced by 1D lenses are expected and especially, at frequencies below a few GHz. Our axisymmetric lens model can account for the observed features of five well sampled ESEs indicating that a cylindrical tube or sphere may be better describing the lens geometry. A systematic search of atypical ESEs at multiple frequencies, from a few hundred MHz to a few GHz, may reveal the geometry of interstellar plasma lenses. 

\section*{Acknowledgements} 
We thank David Kaplan for useful discussion and comments. 
We also thank the anonymous referee and Dr. A. Tuntsov for constructive comments that helped to improve the manuscript.
We acknowledge support from the Research Corporation for Science Advancement’s Scialog program with award ID \#24247. 
We also acknowledge the GBI-NASA monitoring program. The Green Bank Interferometer is a facility of the National Science Foundation operated by the NRAO in support of NASA High Energy Astrophysics programs. MP acknowledges support by the Lyman Jr.~Spitzer Fellowship. 

\bibliographystyle{mnras} 
\bibliography{lens.bib} 

\appendix 
\section{Computation of refracted intensity}\label{app-a}
The computation of the refracted intensity by a 2D axisymmetric plasma lens follows closely that of the 1D lens model presented in \cite{Clegg_1998}, but it is not identical; one needs to consider also the tangential magnification of images and the azimuthal contribution of the extended background sources. 

\begin{figure}
\centering
\includegraphics[width=0.49\textwidth]{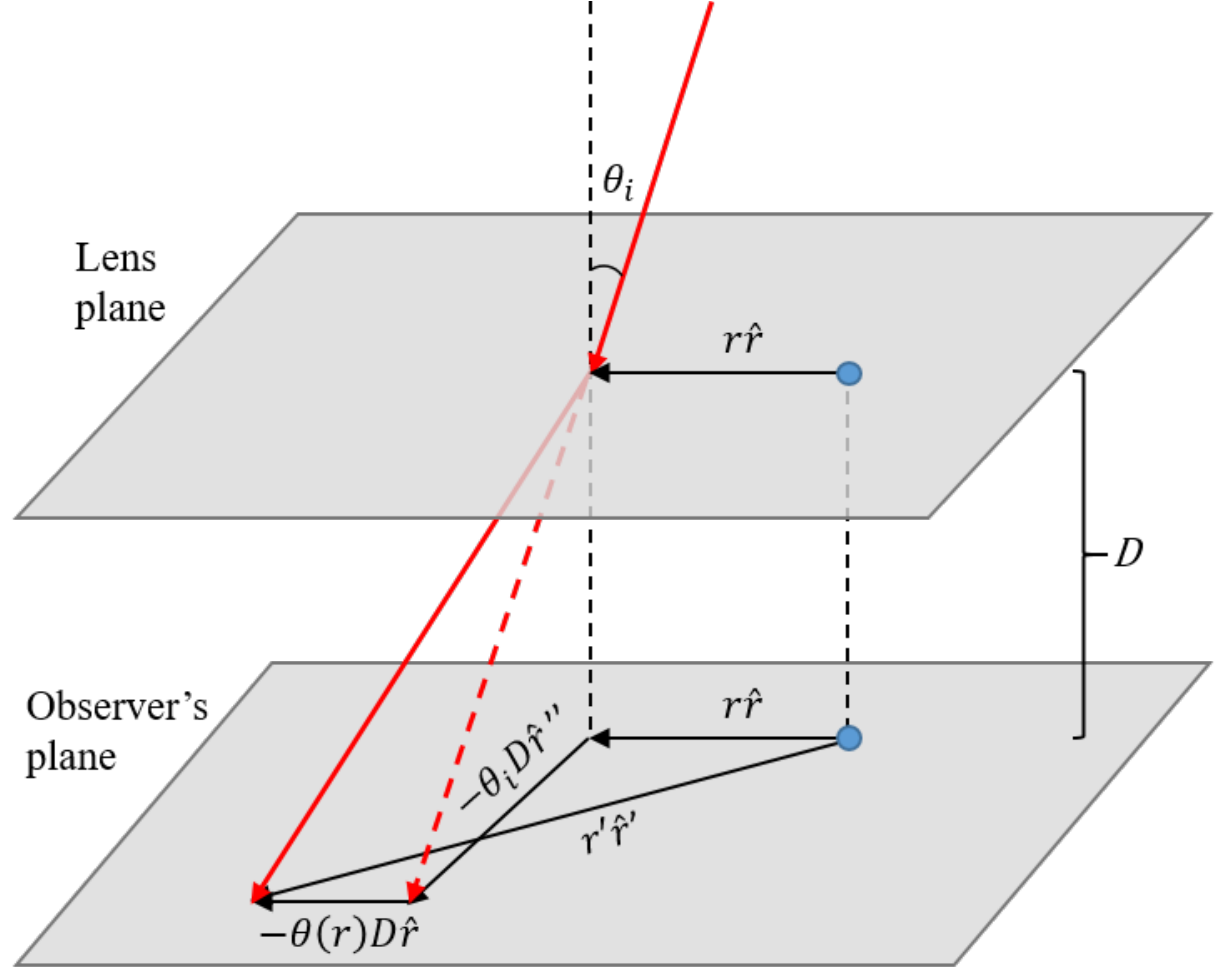}
\caption{A sketch for the ray path geometry in the geometric optics regime. The distance between the lens plane and the observer's plane is denoted as $D$ and is much larger than all other distances displayed on the graph. An arbitrary light ray is shown as an example (red line).The blue dot marks the origin of the coordinate system. The incoming ray hits the lens plane at $r \hat{r}$ with an azimuthal angle $\theta_{\rm i}$. It will be refracted in the $\hat{r}$ direction by an angle $\theta_{\rm r}$, which depends upon the radial distance $r$ and reach the position $r' \hat{r'}$ in the observer's plane. The direction of the non-refracted light ray is also shown (red dashed line).}
\label{fig:raypath}
\end{figure}

Let us consider an axisymmetric lens with a free-electron column density profile described by eq.~(\ref{eq:Ne}). Suppose there is an incoming ray at the lens with an azimuthal angle $\theta_{\rm i}$  and radial direction $\hat{r''}$. Let also $r$ be the radial distance of the point that the incoming ray hits the lens plane. This will be refracted in the $\hat{r}$ direction by an angle $\theta_{\rm r}$, which depends upon the radial distance $r$ and is computed as follows:
\eqb 
\begin{split}
	\theta_{\rm r}(r) & = \frac{\lambda}{2 \pi} \frac{\partial}{\partial r}\Phi(r) \\
                & = - \frac{\lambda^2 r_e N_0}{\pi l^2} r e^{- ( r/l)^2} \\
                & = -  \frac{\alpha}{D} r e^{- ( r/l)^2},
\end{split}
\eqe 
where 
\eqb 
	\Phi(r) = \lambda r_e N_0 \text{ exp } [ -(r/l)^2 ] 
\eqe 
,$\alpha \equiv  \lambda^2 r_e N_0 D/\pi l^2$ is the strength of the lens and $D$ is the distance between the lens plane and the observer's plane. The refracted ray will hit the observer plane at a distance $r'$. A skectch of the geometry is shown in ~\ref{fig:raypath}. In the limit of geometric optics, the path of the light ray is described by:
\eqb
\begin{split}
	r' \hat{r'} & = r \hat{r} - \theta_r(r) D \hat{r} - \theta_{\rm i} D \hat{r''} \\
    r' \hat{r'} & = (r  - \theta_r(r) D) \hat{r} - \theta_{\rm i} D \hat{r''} \\
    r' \hat{r'} & = r ( 1 + \alpha e^{-(r/l) ^2}) \hat{r} - \theta_{\rm i} D \hat{r''} \\
   	u' \hat{r'} & = u ( 1 + \alpha e^{- u^2 }) \hat{r} - \beta_{\rm i} \hat{r''} 
\end{split}
\label{eq:raypath}
\eqe
where 
\eqb 
	u' \equiv \frac{r'}{l} \\
    u \equiv \frac{r}{l} \\
    \beta_{\rm i} \equiv \theta_i \frac{D}{l} = \frac{\theta_{\rm i}}{\theta_l} \\ 
    \theta_l \equiv \frac{l}{D}.
\eqe 
Rearrangement and algebraic manipulation of eq.~(\ref{eq:raypath}) leads to:
\eqb
	u' \hat{r'} + \beta_{\rm i} \hat{r''}  = u ( 1 + \alpha e^{- u^2 }) \hat{r} \\
    u'^2 + 2 \cos{\phi} u' \beta_i + \beta_{\rm i}^2   = u^2 ( 1 + \alpha e^{- u^2 })^2 \\
    u ( 1 + \alpha e^{- u^2 }) - \sqrt{u'^2 + 2 \cos{\phi} u' \beta_{\rm i} + \beta_i^2} = 0
\eqe
where $\cos{\phi} = \hat{r'} \cdot \hat{r''}$ is the angle between the two unit vectors. Upon definition of $\gamma \equiv \sqrt{u'^2 + 2 \cos{\phi} u' \beta_{\rm i} + \beta_{\rm i}^2} $, we find
\eqb 
	u ( 1 + \alpha e^{- u^2 }) - \gamma = 0,
\label{eq:images}
\eqe 
which is similar to eq.~(17) in \cite{Clegg_1998}. 

The source brightness profile can be described by a Gaussian function with zero mean and standard deviation $\sigma \simeq \beta_{\rm s} / 2.355$:
\eqb 
	B(\beta_{\rm i}) = \frac{I_0}{2\pi \sigma^2} e^{ - \frac{\beta_{\rm i}^2}{2\sigma^2}}. 
\eqe 
The refracted light (relative to the unlensed component from the source) at a  point $u'$  from the axisymmetric center  on the observer's plane is given by
\eqb
\begin{split}
    I(u', \alpha) & =\int_{0}^{+4\sigma} \!\!\! {\rm d}\beta_{\rm i}\int_{0}^{2\pi} \!\! {\rm d}\phi \sum_{k = 1}^{n}  B(\beta_{\rm i}) G_{\rm k}(u', \alpha, \beta_{\rm i}, \phi) \beta_i\\
        & =\sum_{i = 1}^{m} \sum_{j = 1}^{l}  \sum_{k = 1}^{n} B(\beta_{\rm i})G_k(u', \alpha, \beta_{\rm i}, \phi) \beta_{\rm i} \Delta\beta_{\rm i} \Delta\phi_{\rm j},
\end{split}
\label{eq:intensity}
\eqe
where $u_{\rm k}$'s are the roots of  eq.~(\ref{eq:images}), $G_{\rm k}$ is the gain factor of the $k$-th refracted image, and $n$ is the total number of images. The radial magnification is given by \citep[eq.~(18)][]{Clegg_1998}:
\eqb
   G_{\rm k, r}= [1 + (1-2u_{\rm k}^2) \alpha e^{-u_{\rm k}^2}]^{-1}
\eqe
and the tangential one is can be written as\citep[eq.~(52)][]{Er_2018}:
\eqb
   G_{\rm k, t}= [1 + \alpha e^{-u_{\rm k}^2}]^{-1}.
   \label{eq:Gkt}
\eqe
The total magnification $G_{\rm k}$ is simply the product of $G_{\rm k, r}$ and $G_{\rm k, t}$. 

The integration step in $\beta_{\rm i}$ used in eq.~(\ref{eq:intensity}) is $\Delta\beta_{\rm i} = 3\sigma / m $, where $m$ is the number of incident rays and depends on $\beta_{\rm s}$. We choose $m\gtrsim 12000 \times \left(\beta_{\rm s}/1\right)$ with an absolute lower bound of 3000 rays. The integration step in the $\phi$  direction used in eq.~(\ref{eq:intensity}) is $\Delta\phi = 2 \pi / l $, where $l$ is the number of azimuthal angles and is fixed to be $45$. The number of possible images is determined by eq.~(\ref{eq:images}) and $n\le 3$, as explained in \cite{Clegg_1998}. 

To compute the refracted intensity maps shown in Fig.~\ref{fig:2-1}, 
we first calculated the light curve of the path with $b = 0$ and then performed a rotation by 2$\pi$.
The temporal resolution of the light curves presented in our paper varies depending on the case, starting from 7 points to 30 points per unit normalized distance (i.e., $d/l$). 
Figs.~\ref{fig:bcr} have the same dimension $90 \times 90$ and are both interpolated from original grids of $9 \times 5$.

\end{document}